\documentclass[onecolumn,amsmath,amssymb]{revtex4-2}
\usepackage{graphicx}
\usepackage{dcolumn}
\usepackage{bm}
\usepackage{amsmath,amssymb,amsfonts}
\usepackage{microtype}
\usepackage{hyperref}
\usepackage{bm}% bold math
\usepackage{tikz}
\usepackage{pgfplots}
\usepgfplotslibrary{groupplots} 
\pgfplotsset{compat=1.18}
\usepackage{booktabs} % 導言區需加入
\usetikzlibrary{calc,3d,angles,quotes,positioning}
\usetikzlibrary{arrows.meta} 
\usepgfplotslibrary{colormaps}
\usepackage{enumitem}

\usepackage{graphicx}

\newcommand{\be}{\begin{equation}}
\newcommand{\ee}{\end{equation}}
\newcommand{\ba}{\begin{eqnarray}}
\newcommand{\ea}{\end{eqnarray}}
\newcommand{\ban}{\begin{eqnarray*}}
\newcommand{\ean}{\end{eqnarray*}}

\begin{document}

%\title{Coherent Control of Anharmonic Molecular Dissociation via Double-Bass Dynamics}
%\title{Coherent Control of Anharmonic Molecular Dissociation Inspired by Double-Bass Dynamics}
\title{Pulse-Shaping for Quantum Dissociation Control of Diatomic Molecules Inspired by Double-Bass Dynamics}
%\title{Bowed-String-Mechanism Inspired Quantum Dissociation Control of Molecules}
%\author{Julian Juhi-Lian Ting\aff{1}}
%\affiliation{\aff{1}De-Font Institute for Advanced Studies, Taichung 40344, Taiwan, R.O.C.}
\author{Julian Juhi-Lian Ting} \email{tingjulian@gmail.com} 
\affiliation{De-Font Institute for Advanced Studies, Taichung 40344, Taiwan, R.O.C.}

%\authordeclaration{There are no competing interests.}

\begin{abstract}

Drawing on the acoustic mechanics of accelerating Helmholtz motion in bowed strings, 
we present a pulse-shaping strategy to steer the multiphoton dissociation of vibrationally excited diatomic molecules. 
Under intense infrared fields, molecules often exhibit unpredictable control windows caused by the chaotic fragmentation of critical dissociation thresholds. 
By implementing a harmonic bowing-analogue transient envelope governed by a $\mathcal{C}^1$ continuity condition 
and vanishing initial boundary conditions ($T(0)=0, T'(0)=0$), 
we demonstrate that non-linear front-flank acceleration collapses the multi-valued threshold topology 
into a predictable, single-valued dissociation corridor. 
Quantum phase-space Wigner diagnostics confirm that this engineered trajectory dilates microscopic cantori bottlenecks and suppresses wave-packet fragmentation. 
%Numerical simulations further verify that spatial wave-packet localization is preserved even under Markovian thermal dephasing and rotational manifold detuning. 
Numerical simulations further verify that spatial wave-packet localization is preserved even under Markovian thermal dephasing.
Comparative evaluations against iterative optimal control protocols and conventional synthesized two-color Gaussian envelopes demonstrate that standard smooth pulses fail to arrest non-hyperbolic cantori stickiness due to insufficient sub-picosecond front-flank acceleration. 
In contrast, the boundary-constrained harmonic shaper enforces a robust single-valued dissociation corridor within an ultra-short interaction window of $1.5\text{ ps}$.

\end{abstract}

\date{\today}
\maketitle
\raggedbottom

%Cover letter: In 1995 I published a paper on modeling of NO dissociation using Morse potential. Here is a follow-up. One of the interesting point is that I got inspiration from a double-bass player who conducted research about how to start an oscillation of a double-bass quickly, which is isomorphic to the diatomic oscillator I considered.
%Please do not arbitrarily decline my paper which tresspasing the peer review system and is a scieitific misconduct. If you are qualified as a referee you can appoint yourslef as a referee, your opinion count as one, 
%but you are not a superreferee. As for your comment saying my present has not enough novety from the previous one is totally incorrect.
%You simply need to take out both paper and compare everything become clear. I have also added a little explaination at the second paragraph. 

%Cover letter: This is one of the most interesting paper I have ever wrote. Firstly, the inspiration of Guettler is interesting and rare. I not only conducted numerical calculations but also verified using Wigner's distribution. Furthermore, I compared with others' work. My previous work on NO was done during my PHD years. Now I can see how much I have progressed over the years.

\section{Introduction}

The physical principles governing vibrations in molecular systems share intriguing parallels with those of string instruments~\citep{Weinreich1993}. 
While the molecular community is often considered theoretically advanced, 
musical acoustics occasionally uncovers unique dynamical principles that remain underappreciated in quantum control. 
Among bowed-string instruments, 
the double bass occupies a unique position due to its exceptionally low fundamental frequencies and resulting high acoustic inertia. 
This characteristic poses a severe challenge for the performer, 
who must establish a clean sound attack within fractions of a second when the orchestra engages in rapid \textit{allegro} or \textit{presto} passages. 
This musical demand for rapid acoustic onset precisely mirrors a critical challenge in molecular engineering. 
In particular, Guettler's methods for managing transient boundary conditions are designed to accelerate steady-state Helmholtz motion while suppressing chaotic stick-slip noise~\citep{Guettler1997, Guettler2002, Guettler2012,Galluzzo2014}, 
which directly parallel the requirements for steering highly excited quantum states. 
Specifically, in the multiphoton dissociation of anharmonic molecules such as nitrogen oxide (NO), 
optimal control fields must initiate under explicit soft-starting constraints to avoid spurious non-adiabatic effects~\citep{Ting1992, Ting1994, Ting1995}, 
while concurrently driving the system past phase-space bottlenecks with sub-picosecond acceleration.

%We considered the mininum fields required to achieving the dissociation of nitrogen oxide (NO) molecules years ago~\citep{Ting1992a,Ting1994,Ting1995},
%and noticed the driving function should start softly whereas the critical laser intensity required for infrared multiphoton dissociation exhibits a fragmented and multi-valued topology up to state $n=32$,
%without considering the initial transit periods. %(about $60$ cycles or $1.5 ns$, see Figure \ref{fig:dynamics_suppression} below).
%While earlier investigations into the multiphoton dissociation of NO focused on determining minimum field intensities 
%over longer durations without considering the initial transit periods, 
%this work emphasizes the critical role of the sub-picosecond 'Attack' phase. 
%We demonstrate that by engineering the pulse over a total control window of $1.5 ns$, 
%the deterministic shaping of the initial transient,
%covering approximately $60$ optical cycles,
%is sufficient to bypass the chaotic fragmentation that plagued previous sudden-on models.

%The rovibrational degrees of freedom in diatomic molecules complicate state engineering within deeply embedded vibrational manifolds ($n = 25$ to $32$)~\citep{DeMille2015}. 
The vibrational degrees of freedom in diatomic molecules complicate state engineering within deeply embedded vibrational manifolds ($n = 25$ to $32$)~\citep{DeMille2015}.
In this intermediate multi-photon resonance region, which sits at the onset of the hard-chaos threshold of the driven Morse potential, 
the critical field intensity required for multiphoton dissociation exhibits a fragmented, multi-valued topology near resonance. 
This complex topography originates from underlying phase-space bottlenecks where subharmonic resonances, wave-packet delocalization, 
and bond-softening induce non-hyperbolic chaotic scattering~\citep{Constantoudis2001}. 
To navigate these topological complexities, closed-loop adaptive learning networks governed by genetic algorithms have been used 
to optimize optical fields based on empirical fitness feedback ~\citep{Judson1992}. 
However, without external statistical filtering, these optimization processes frequently yield complex fields that are highly susceptible to local optima, 
as they cannot explicitly account for the complex phase-space boundaries inherent in strongly driven anharmonic oscillators ~\citep{Bucksbaum2025}.

To resolve this threshold fragmentation, 
we introduce a pulse-shaping strategy derived from the transient mechanics of bowed-string instruments. 
In bowed-string dynamics, particularly under the demanding low-frequency of double bass performance, 
triggering steady-state periodic Helmholtz motion within acceptable auditory transient limits requires a balance of initial bowing parameters, 
a criterion established by Guettler ~\citep{Guettler1997,Guettler2002,Guettler2012} which,
by achieving rapid onset, bypasses prolonged, chaotic stick-slip transients ~\citep{Woodhouse2004,Galluzzo2014,Makarenkov2017} 
through strategic over-driving bow acceleration, see $A_{over}$ in Eq. ~\eqref{bowing},
and static pre-force alignment given by $\varepsilon_d$ in Eq.~ \eqref{two}. 
Furthermore, this cross-scale control efficacy exploits the system's dynamic memory of inertia, 
which manifests in molecular dynamics as constructive coherent phase-overlapping between the multi-color fields ({\it vide infra}, Eq. \eqref{phaseoverlapping}).

In this work, we show that the acoustic mechanics of string instrument bowing, originally identified for suppressing pre-Helmholtz friction noise, can be translated into an effective quantum control protocol to smooth the dissociation landscape of diatomic molecules. 
We investigate a vibrating NO molecule interacting with an intense two-color laser field consisting of a fundamental driving frequency ($\Omega$) and its third harmonic ($3\Omega$). 
Rather than adopting standard rectangular, adiabatic, or unconstrained continuous envelopes, we introduce a customized time-dependent pulse shaper that reproduces the non-linear acceleration trajectory of a bow attack. 
By solving the time-dependent Schr\"odinger equation via a split-operator Fast Fourier Transform (FFT) grid method with infinite-wall boundary conditions~\citep{Alhendi2005}, we systematically evaluate how the initial front-flank curvature and peak-overdrive amplitude govern the fragmentation of the critical dissociation threshold ($A_c$). 
While contemporary coherent control frameworks rely predominantly on standard Gaussian envelopes or transform-limited profiles, 
recent studies in resonant nonlinear interactions have confirmed that conventional smooth pulses are inherently suboptimal for steering systems governed by strong anharmonicity and phase-space dispersion~\citep{Meron2025}. 
Without temporal derivative constraints at the activation boundary layers, smooth envelopes cannot outpace the power-law trapping associated with non-hyperbolic chaotic scattering.
By engineering the initial transient over a total interaction window of $1.5\text{ ps}$ (approximately 60 optical cycles), 
this deterministic boundary-layer formulation collapses the chaotic multi-valued thresholds into a single-valued dissociation corridor, 
establishing topological stabilization against phase-space stickiness.

The paper is structured as follows. 
Section~\ref{model} establishes the theoretical model of the modulated Hamiltonian and formulates the harmonic bowing-analogue control function with its boundary constraints. 
%Section~\ref{numerical} presents the numerical results, detailing the collapse of multiple thresholds, phase-space Wigner diagnostics, robustness against environmental decoherence and rotational detuning, as well as comparative evaluations against optimal control protocols and synthesized two-color Gaussian envelopes. 
Section~\ref{numerical} presents the numerical results, detailing the collapse of multiple thresholds, phase-space Wigner diagnostics, robustness against environmental decoherence, as well as comparative evaluations against optimal control protocols and synthesized two-color Gaussian envelopes. 
Finally, Section~\ref{conclusion} summarizes the main findings and provides concluding remarks.

\section{Theoretical Model and Bow-String Analogue Operators}
\label{model}

\subsection{The Total Modulated Hamiltonian and Molecular Model}
\label{sec:hamiltonian}

To evaluate the coherent transient dynamics of a diatomic molecule under intense laser fields, 
we construct a fully coupled non-linear Hamiltonian within the semiclassical dipole approximation~\citep{Ting1992,Ting1994,Ting1995, Geng2002a, Varro2023}. 
The one-dimensional non-rotating nuclear vibrational dynamics of a system such as $\text{NO}(X\,^2\Pi)$, subjected to a static electric field $\varepsilon_d$ and a pulse-shaped laser field $E(t)$, is governed by:
\begin{equation}
\hat{H}(t) = -\frac{\hbar^2}{2m}\frac{d^2}{dx^2} + V_M(x) - \mu(x, E(t))\cdot E(t) - \mu_1 x \varepsilon_d \;,
\label{two}
\end{equation}
where $m$ is the reduced mass of the molecule, and $x = r - r_0$ represents the nuclear displacement from the equilibrium internuclear distance $r_0$. The unperturbed field-free molecular potential is modeled via the empirical Morse potential:
\begin{equation}
V_M(x) = D \left( 1 - e^{-\alpha x} \right)^2 \;,
\label{eq:morse}
\end{equation}
where $D$ denotes the unperturbed dissociation energy and $\alpha$ dictates the spatial anharmonicity. 
These Morse parameters ($D=0.2426$~a.u., $\alpha=1.450$~a.u.$^{-1}$) are derived from the experimental spectroscopic constants ($\omega_e = 1904.2\,\text{cm}^{-1}$ and $\omega_e x_e = 14.1\,\text{cm}^{-1}$)~\citep{Huber1979}. For low-to-intermediate vibrational manifolds ($n \le 20$), Eq.~\eqref{eq:morse} closely reproduces the empirical Rydberg-Klein-Rees (RKR) potential energy surface~\citep{Gilmore1965}. Although ab initio surfaces exhibit long-range van der Waals tails at highly stretched configurations ($n \ge 32$), the Morse potential provides a robust analytical baseline that captures the essential spatial anharmonicity and hard-chaos onset required to investigate threshold collapse. The resulting Morse shape factor $s = \sqrt{2mD}/(\alpha\hbar) \approx 54.54$ accommodates 55 bound vibrational states ($n = 0, \dots, 54$).

To incorporate electronic coordinate distortions in highly stretched vibrational states ($n \le 32$), the generalized coordinate-dependent dipole moment function $\mu(x, E(t))$ is truncated to first order in the nuclear displacement and polarizability gradient~\citep{Varro2023}:
\begin{equation}
\mu(x, E(t)) = \mu_0 + \mu_1 x + \alpha_0 E(t) + \alpha_1 x E(t) \;,
\label{dipole}
\end{equation}
where $\mu_0$ is the permanent dipole moment, $\mu_1 = (\partial \mu / \partial x)_{x=0}$ is its linear gradient, $\alpha_0$ is the static electronic polarizability, and $\alpha_1 = (\partial \alpha_{\text{e}} / \partial x)_{x=0}$ is the polarizability derivative. All physical constants and parameters for the NO system are summarized in Table~\ref{tab:parameters}.

When peak laser intensities exceed $\sim 10^{12}\,\text{W/cm}^2$, non-adiabatic coupling between electronic and nuclear motions can trigger multiphoton ionization or non-resonant electronic transitions, necessitating a fully coupled electron--nuclear time-dependent Schr\"odinger equation (TDSE)~\citep{Ghosh2024}. In our protocol, the peak field amplitude is strictly constrained below this non-adiabatic threshold~\citep{Tiwari2008}, ensuring the physical validity of the single-surface nuclear vibrational description. Substituting Eq.~\eqref{dipole} into Eq.~\eqref{two} and collecting the electrostatic interaction terms yields the effective modulated potential surface $V_m(x)$:
\begin{equation}
V_m(x) = D \left( 1 - e^{-\alpha x} \right)^2 - \left[ \mu_1 \varepsilon_d + \mu_1 E(t) + \alpha_1 E^2(t) \right] x - \mu_0 E(t) - \alpha_0 E^2(t) \;.
\label{eq:modulated_potential}
\end{equation}
The linear coefficients inside the brackets parameterize the dynamic, coordinate-dependent Stark shift experienced by the vibrational manifold. 

The present single-surface potential deformation is different from multi-surface laser-dressed potential energy curves (DPECs) commonly observed in intense-field photodissociation dynamics~\citep{Dey2016}. 
In strong-field regimes involving multiple electronic manifolds, laser dressing generates complex topologies featuring light-induced conical intersections (LICIs) and avoided crossings. In contrast, our framework operates strictly within the non-ionizing single-surface Born-Oppenheimer regime ($I < 10^{12}\,\text{W/cm}^2$), where non-adiabatic electronic state mixing is absent. The field-induced profile $V_m(x)$ depicted in Fig.~\ref{fig:pot_and_pulse}(a) represents the instantaneous, coordinate-dependent Stark-shifted effective potential on the ground $\text{NO}(X\,^2\Pi)$ surface, providing a clear visual representation of outer-barrier depression under static electrostatic biasing.

The physical mechanism of this electrostatic biasing can be understood through an acoustic analogy. 
In macroscopic string mechanics, a plucked string (pizzicato) merely releases a symmetric spatial deformation that decays anisotropically without sustaining a limit cycle. In contrast, bowing introduces a directional velocity and frictional force that explicitly breaks spatial and temporal symmetry, sustaining asymmetric Helmholtz motion. Analogously, an unperturbed symmetric laser field induces isotropic wave-packet dispersion. Introducing the static electric field $\varepsilon_d$ and the asymmetric dual-color driver ($\Omega + 3\Omega$) acts as a fundamental symmetry-breaking agent. 

This field-induced bias depresses the outer Morse barrier along $x \to \infty$ [as illustrated in Fig.~\ref{fig:pot_and_pulse}(a)], breaking the spatial symmetry of the potential and guiding the wave-packet deterministically into the primary acceleration corridor. The dimensionless static field parameter $K = \mu_1 \varepsilon_d / (\alpha D)$ quantifies this static electrostatic bias relative to the unperturbed dissociation energy $D$. Much like a double-bass player adjusting the static contact force and bow angle to bias the frictional stick-slip threshold, this pre-exposure bias depresses the binding barrier, tailoring the bound-state level spacings to optimize the initial wave-packet projection into the single-valued dissociation corridor.

\begin{table*}[h!]
\centering
\caption{Physical constants and model parameters for the $\text{NO}(X\,^2\Pi)$ system and the bowing-analogue laser pulse.}
\label{tab:parameters}
\begin{tabular}{lll}
\hline\hline
Parameter & Description & Value \\
\hline
$D$ & Morse dissociation energy & $0.2426\text{ a.u.}$ ($6.60\text{ eV}$) \\
$\alpha$ & Morse spatial range parameter & $1.450\text{ a.u.}^{-1}$ ($2.74\text{ \AA}^{-1}$) \\
$m$ & Reduced mass of NO & $13612\text{ a.u.}$ ($7.466\text{ amu}$) \\
$r_0$ & Equilibrium bond distance & $2.175\text{ a.u.}$ ($1.151\text{ \AA}$) \\
$\omega_0$ & Fundamental vibrational frequency & $0.00863\text{ a.u.}$ ($1894\text{ cm}^{-1}$) \\
$\mu_0$ & Permanent dipole moment & $0.059\text{ a.u.}$ ($0.15\text{ D}$) \\
$\mu_1$ & Dipole derivative $(\partial\mu/\partial x)_{x=0}$ & $0.50\text{ a.u.}$ \\
$\alpha_0$ & Static electronic polarizability & $11.5\text{ a.u.}$ \\
$\alpha_1$ & Polarizability derivative $(\partial\alpha_e/\partial x)_{x=0}$ & $7.00\text{ a.u.}$ \\
$K$ & Dimensionless static-bias coupling & $0.02$ \\
$A_{\text{over}}$ & Bowing peak-overdrive ratio & $1.35$ \\
$\gamma$ & Front-flank acceleration exponent & $1.60$ \\
$\tau_{\text{att}}$ & Attack transient duration & $0.25\text{ ps}$ \\
$\tau_{\text{plateau}}$ & Relaxation to plateau threshold & $0.45\text{ ps}$ \\
$\tau_{\text{fall}}$ & Onset of field release & $1.20\text{ ps}$ \\
$\tau_{\text{total}}$ & Total interaction window & $1.50\text{ ps}$ \\
$A_2/A_1$ & Ratio of third harmonic to fundamental & 0.2 \\
\hline\hline
\end{tabular}
\end{table*}
 
%The reasons to choose $\Omega$ and $3\Omega$ are: Firstly, due to symmetry breaking in both time and space consideration.
%For 1D Morse potential, the direction of dissociation is in $x \to \infty$. Using space asymmetryc laser field we can push
%the molecule in that direction.
%The second reason is due to quantum path interference. For a molecule to jump to a higher energy level,
%there are two paths. The first one is absorpting three photons of frequency $\Omega$ and the second possibility is to absorp one
%photon with frequency $3 \Omega$. They have the same energy. But these two paths will interfere.
%The third reason is because third-harmonic is a common practice in experiment.

\subsection{The Bowing-Analogue Laser Control Function}
\label{analogy}

To actively steer the wave-packet away from the non-hyperbolic chaotic stickiness maps and 
suppress the fragmentation of multiple dissociation thresholds, 
we introduce a two-color laser field profile $E(t)$ as before~\citep{Ting1995}. 
The field consists of a fundamental driving frequency $\Omega$ combined with its third harmonic $3\Omega$:
\begin{equation}
E(t) = T(t) \cdot \left[ A_1 \sin(\Omega t) + A_2 \sin(3\Omega t + \phi) \right] \;,
\label{phaseoverlapping}
\end{equation}
where $A_1$ and $A_2$ represent the peak field amplitudes, and $\phi$ defines the relative phase difference between the two colors. In all numerical calculations, the amplitude ratio between the third harmonic and the fundamental frequency is fixed at $A_2 / A_1 = 0.2$, with $A_1$ dynamically scaled to locate the critical dissociation threshold $A_c$.
The rationale for selecting the $\Omega$ and $3\Omega$ dual-color combination is threefold. 
First, it induces both temporal and spatial symmetry breaking. 
For a one-dimensional Morse potential, molecular dissociation proceeds exclusively along the $x \to \infty$ direction. 
By using a spatially asymmetric laser field, the wave-packet can be deterministically driven toward the dissociation corridor. 
Second, this configuration exploits quantum path interference. 
Transitioning the wave-packet to a higher vibrational level can occur via two coherent pathways: 
the simultaneous absorption of three photons of frequency $\Omega$, or the absorption of a single photon of frequency $3\Omega$. 
Because these degenerate pathways share the same target energy, their quantum interference can be actively controlled via the relative phase $\phi$, 
as demonstrated in synthesized two-color intense laser control schemes~\citep{Ghosh2024}. 
Thirdly, third-harmonic generation is a well-established and robust practice in modern ultrafast laser experiments.

\subsection{Harmonic Bowing Shaper and $\mathcal{C}^{1}$ Continuity Constraints}
\label{sec:bowing_shaper}

Rather than adopting conventional adiabatic or sudden-on rectangular pulse envelopes, the time-dependent envelope operator $T(t)$ modulates the composite laser field architecture and is engineered to reflect the physical acceleration trajectory of a bow stroke during its attack transient phase. In classical string acoustics, Guettler's criterion indicates that establishing stable, periodic Helmholtz motion while suppressing chaotic stick-slip transients requires a critical combination of rapid bow-force alignment and non-linear bow-velocity acceleration~\citep{Guettler1997, Guettler2002}. Traditional continuous control envelopes, such as standard Gaussian or adiabatic multi-harmonic waveforms, fail to deliver sufficient front-flank acceleration within sub-picosecond scales, 
thereby leaving the molecular wave packet trapped inside non-hyperbolic chaotic cantori structures. 

\begin{figure}[htbp]
\centering
\begin{tikzpicture}

% ==================== SUBFIGURE (a): POTENTIAL DISTORTION ====================
\begin{axis}[
    name=potential,
    width=0.48\textwidth,
    height=6cm,
    xmin=-0.8, xmax=4.5,
    ymin=-0.1, ymax=0.4,
    xlabel={$x$ (a.u.)},
    ylabel={Potential Energy (a.u.)},
    title={(a) Potential Energy Surface Distortion},
    legend pos=north east,
    legend style={draw=none, font=\small},
    xtick={0, 1.0, 2.0, 3.0, 4.0},
    ytick={0, 0.1, 0.2, 0.3},
    tick label style={font=\small},
    axis lines=left
]
    % Field-Free Morse Potential: D*(1-exp(-\alpha*x))^2  (D=0.2426, \alpha=1.45)
    \addplot[domain=-0.5:4.2, samples=100, thick, black] {0.2426*(1-exp(-1.45*x))^2};
    \addlegendentry{Field-Free $V_M(x)$}
    
    % Field-Induced Modulated Potential (Tilted downward at large x)
    \addplot[domain=-0.5:4.2, samples=100, dashed, red, thick] {0.2426*(1-exp(-1.45*x))^2 - 0.035*x};
    \addlegendentry{Field-Induced $V_m(x)$}
    
    % Barrier lowering arrow
    \draw[->, >=stealth, thick, blue] (axis cs:2.5, 0.23) -- (axis cs:2.5, 0.14) 
        node[midway, right, font=\small] {Barrier Lowering};
\end{axis}

% ==================== SUBFIGURE (b): PULSE SHAPE ====================
\begin{axis}[
    at={(potential.right of south east)},
    anchor=left of south west,
    xshift=1.5cm,
    width=0.48\textwidth,
    height=6cm,
    xmin=0, xmax=1.6,
    ymin=0, ymax=1.5,
    xlabel={Time $t$ (ps)},
    ylabel={Envelope $T(t)$},
    title={(b) Bowing-Analogue Pulse Envelope},
    axis lines=left,
    xtick={0, 0.25, 0.45, 1.2, 1.5},
    xticklabels={$0$, $\tau_{\text{att}}$, $\tau_{\text{plateau}}$, $\tau_{\text{fall}}$, $\tau_{\text{total}}$},
    ytick={0, 1.0, 1.35},
    yticklabels={$0$, $1$, $A_{\text{over}}$},
    tick label style={font=\small}
]
    % Phase 1: Attack Transient (0 to \tau_att=0.25) - Nonlinear Acceleration
    \addplot[domain=0:0.25, samples=50, ultra thick, red] {1.35 * sin(deg(pi*x/(2*0.25)))^1.6};
    
    % Phase 2: Relaxation to Plateau (\tau_att to \tau_plateau=0.45)
    \addplot[domain=0.25:0.45, samples=50, ultra thick, red] {1 + (1.35 - 1) * cos(deg(pi*(x-0.25)/(2*(0.45-0.25))))^2};
    
    % Phase 3: Steady-State Plateau (\tau_plateau to \tau_fall=1.2)
    \addplot[domain=0.45:1.2, samples=10, ultra thick, red] {1.0};
    
    % Phase 4: Field Release / Decay (\tau_fall to \tau_total=1.5)
    \addplot[domain=1.2:1.5, samples=50, ultra thick, red] {cos(deg(pi*(x-1.2)/(2*(1.5-1.2))))^2};
    
    % Auxiliary lines for clarity
    \draw[dashed, gray] (axis cs:0.25,0) -- (axis cs:0.25,1.35);
    \draw[dashed, gray] (axis cs:0.45,0) -- (axis cs:0.45,1.0);
    \draw[dashed, gray] (axis cs:1.2,0) -- (axis cs:1.2,1.0);
    \draw[dashed, gray] (axis cs:0,1.35) -- (axis cs:0.25,1.35);
    \draw[dashed, gray] (axis cs:0,1.0) -- (axis cs:0.45,1.0);
\end{axis}

\end{tikzpicture}
\caption{(a) Comparison between the unperturbed field-free Morse potential $V_M(x)$ (solid black curve) and the field-induced modulated potential surface $V_m(x)$ showing the Stark-induced barrier depression. 
(b) Time-dependent profile of the piecewise continuous Bowing Shaper $T(t)$, highlighting the non-linear front-flank peak overdrive ($A_{\text{over}}$) and smooth $C^1$ switching nodes.}
\label{fig:pot_and_pulse}
\end{figure}
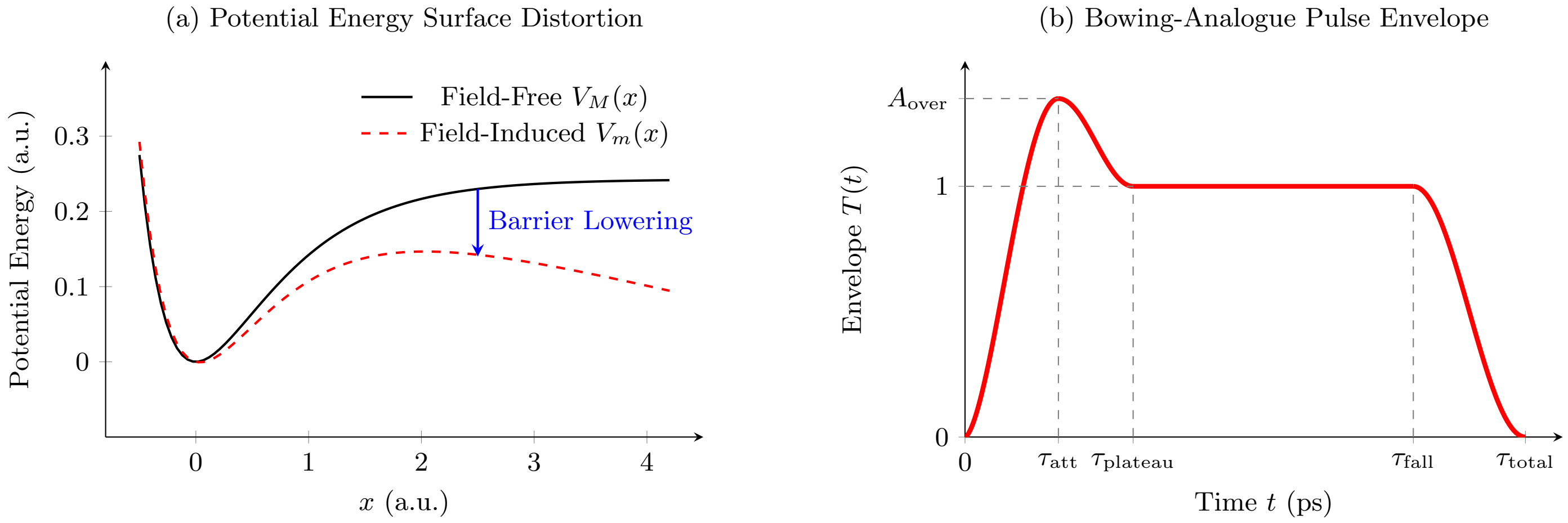

To overcome this limitation and enforce a strict physical constraint that mirrors a double-bass bow attack without introducing non-physical acceleration kinks, 
we define the Harmonic Bowing Shaper as a piecewise harmonic continuous function:
\begin{equation}
T(t) = \begin{cases} 
A_{\text{over}} \sin^\gamma \left( \dfrac{\pi t}{2\tau_{\text{att}}} \right), & 0 \le t < \tau_{\text{att}} \\[10pt] 
1 + (A_{\text{over}} - 1) \cos^2 \left[ \dfrac{\pi(t - \tau_{\text{att}})}{2(\tau_{\text{plateau}} - \tau_{\text{att}})} \right], & \tau_{\text{att}} \le t < \tau_{\text{plateau}} \\[10pt] 
1, & \tau_{\text{plateau}} \le t < \tau_{\text{fall}} \\[10pt] 
\cos^2 \left[ \dfrac{\pi(t - \tau_{\text{fall}})}{2\tau_{\text{decay}}} \right], & \tau_{\text{fall}} \le t \le \tau_{\text{total}} 
\end{cases}
\label{bowing}
\end{equation}
where $\gamma > 1.0$ is the acceleration exponent, $A_{\text{over}} \ge 1.0$ denotes the peak-overdrive amplitude ratio, and $\tau_{\text{att}}$, $\tau_{\text{plateau}}$, and $\tau_{\text{fall}}$ represent the absolute temporal switching coordinates marking the termination of the front-flank attack, the completion of relaxation into the steady-state plateau corridor, and the onset of field release, respectively. The trailing decay duration is defined by $\tau_{\text{decay}} = \tau_{\text{total}} - \tau_{\text{fall}}$, satisfying the total interaction duration constraint $\tau_{\text{total}} = 1.50\,\text{ps}$. The complete temporal trajectory of this piecewise shaper is explicitly mapped out in Fig.~\ref{fig:pot_and_pulse}(b), 
illustrating how the pulse profile achieves acceleration while smoothly transitioning across the control nodes.

To eliminate non-physical high-frequency Fourier sideband noise and guarantee the stability of the Suzuki-Trotter numerical propagation across the highly stretched bound states ($n \le 32$), the piecewise boundaries of Eq.~\eqref{bowing} are formulated to preserve both $\mathcal{C}^0$ and $\mathcal{C}^1$ continuity across all switching nodes by construction. At the initial activation $t=0$, enforcing $\gamma > 1.0$ ensures a vanishing temporal derivative ($dT/dt|_{t=0} = 0$), providing a microscopic soft start that suppresses instantaneous switching oscillations. At the primary synchronization node $t = \tau_{\text{att}}$, matching the boundary limits $\lim_{t \to \tau_{\text{att}}^{-}} T(t) = \lim_{t \to \tau_{\text{att}}^{+}} T(t) = A_{\text{over}}$ establishes local $\mathcal{C}^0$ continuity. Concurrently, evaluating the first-order temporal derivatives confirms that both left-hand and right-hand gradients vanish simultaneously:
\begin{equation}
\lim_{t \to \tau_{\text{att}}^{-}} \frac{dT}{dt} = \lim_{t \to \tau_{\text{att}}^{+}} \frac{dT}{dt} = 0 \;.
\label{eq:C1_matching}
\end{equation}
Likewise, across the plateau transitions at $t = \tau_{\text{plateau}}$ and $t = \tau_{\text{fall}}$, the function smoothly matches unity ($\lim T(t) = 1$) and the temporal derivatives vanish identically ($\lim dT/dt = 0$), 
avoiding any artificial sharp acceleration discontinuities that could induce grid propagation artifacts.

This control architecture explicitly encapsulates two high-level strategies adapted from Guettler's bowing strategy, 
which are here generalized to a dual-color $\Omega + 3\Omega$ laser field configuration. 
First, it employs the protocol of \textit{preferential over-acceleration}; 
by engineering the transient peak overdrive ratio to $A_{\text{over}} \ge 1.0$, 
the laser field rapidly drives the system past chaotic phase-space barriers before quantum wave-packet delocalization can trap the trajectories. 
Second, this fast-vamp envelope operates in tandem with the dynamic inertial memory of the vibrating molecular bond. 
By locking the relative phase between the fundamental driving frequency $\Omega$ and its third harmonic to $\phi = \pi$, the $3\Omega$ pulse component injects a localized energy flux precisely when the wave packet reaches its classical turning point ($x > 0$). This results in a constructive coherent superposition that utilizes the system's intrinsic phase history to maximize the directional dissociation flux along the linear corridor.

\subsection{Asymptotic Boundary Behavior and Switching Transients}

To verify that the initial profile of Eq.~\eqref{bowing} fulfills the formal requirements of a macroscopic soft-starting condition, 
we analyze the asymptotic boundary behavior of the envelope function $T(t)$ and its temporal derivatives at the initial activation point $t \to 0^{+}$. 
Evaluating the function value under the non-linear front-flank acceleration regime where the exponent is rigidly bounded via $\gamma > 1.0$, 
the initial constraint collapses to zero:
\begin{equation}
\lim_{t \to 0^{+}} T(t) = \lim_{t \to 0^{+}} A_{\text{over}} \sin^\gamma \left( \frac{\pi t}{2\tau_{\text{att}}} \right) = 0 \;.
\label{eq:soft_start_C0}
\end{equation}
Performing a temporal differentiation of $T(t)$ in Eq.~\eqref{bowing} isolates the microscopic boundary derivatives evaluating the initial velocity and acceleration limits of the envelope. 
For the first-order derivative mapping the pulse velocity, the boundary response evaluates to:
\begin{equation}
\left. \frac{dT}{dt} \right|_{t=0^{+}} = \lim_{t \to 0^{+}} \frac{\gamma \pi A_{\text{over}}}{2\tau_{\text{att}}} \cos\left( \frac{\pi t}{2\tau_{\text{att}}} \right) \sin^{\gamma - 1}\left( \frac{\pi t}{2\tau_{\text{att}}} \right) = 0 \;,
\label{eq:soft_start_C1}
\end{equation}
which identically vanishes for any acceleration profile satisfying $\gamma > 1.0$. 
This analytic boundary-layer formulation rectifies the limitations of sudden-on expressions by ensuring that temporal derivatives do not fracture the phase-space boundaries at the initial activation point. 
Under an abrupt step-function turn-on, non-adiabatic switching perturbations persist for approximately 23 optical cycles, which severely destabilizes initial wave-packet coherence. 
By locking both the envelope values and the temporal gradients to zero at the initial boundary layer, 
the vanishing boundary conditions of $T(0)=0$ and $T'(0)=0$ effectively extinguish this 23-cycle switching transient at its source, 
preventing the initial phase-space fracturing inherent in conventional sudden-on perturbations.

Experimentally, sub-femtosecond relative phase locking and coherent waveform synthesis between a fundamental field ($\Omega$) and 
its third harmonic ($3\Omega$) have become routinely accessible in ultrafast optics via active carrier-envelope phase stabilization networks, 
inline dichroic delay lines, or liquid-crystal spatial light modulators (SLMs). 
In particular, recent advances in Fourier-plane 4f pulse-shaping systems have demonstrated that arbitrary analytical phase profiles and boundary-engineered waveforms can be experimentally imprinted onto sub-10 fs pulses with high fidelity to manipulate resonant non-linear interactions~\citep{Meron2025}. 
These mature pulse-shaping technologies ensure that the temporal synchronization and the front-flank acceleration profile required by Eq.~\eqref{bowing} can be reliably synthesized in modern laboratory environments.

By simultaneously satisfying the $\mathcal{C}^1$ continuity condition in Eq.~\eqref{eq:C1_matching} 
to eliminate gradient mismatches and locking the relative phase at $\phi = \pi$ at the classical vibrational turning point ($x > 0$), 
the dynamic envelope ensures smooth energy injection. 
This phase-locked synchronization delivers localized coherent energy at an expansion rate that outruns the power-law decay of non-hyperbolic chaotic scattering, 
preserving the wave-packet from subharmonic resonance zones and preventing phase-space fragmentation.

\subsection{Quantum Phase-Space Diagnostics: The Wigner Operator}

To track non-classical state splitting and diagnose whether the transient shaper successfully prevents wave-packet delocalization, 
we monitor the time evolution of the system in phase space using the Wigner quasiprobability distribution. 
Adopting the dimensionless grid transformation~\citep{Foldi2002}, 
the time-dependent quantum state $|\psi(t)\rangle$ obtained from the split-operator FFT grid propagation is mapped onto the 
coordinate-momentum phase space $(x, p)$ via:
\begin{equation}
W(x, p, t) = \frac{1}{2\pi\hbar} \int_{-\infty}^{+\infty} \langle \psi(t) \mid x + \frac{u}{2} \rangle \langle x - \frac{u}{2} \mid \psi(t) \rangle e^{i p u / \hbar} du \;,
\label{wigner}
\end{equation}
where $p$ represents the classical conjugate momentum, and $u$ serves as the spatial coordinate displacement integration variable.

To quantify the capability of the bowing shaper to suppress the spontaneous formation of fragmented Schr\"odinger-cat states, 
we compute the global non-classicality index $M_{\text{nc}}(t)$:
\begin{equation}
M_{\text{nc}}(t) = 1 - \frac{|I_+(t)| - |I_-(t)|}{|I_+(t)| + |I_-(t)|} \;,
\end{equation}
where $I_+(t)$ and $I_-(t)$ represent the phase-space integrals of $W(x, p, t)$ restricted to the strictly positive and strictly negative volumes of the distribution, respectively:
\begin{equation}
\begin{split}
I_{\pm}(t) = & \iint_{\mathcal{D}_{\pm}} W(x, p, t) \, dx \, dp \; , \\
\mathcal{D}_{\pm} = & \{ (x,p) \mid \text{sgn}[W(x,p,t)] = \pm 1 \} \;.
\end{split}
\label{volume}
\end{equation}

A clean, classical-like single-peaked wave packet corresponds to $M_{\text{nc}} \to 0$, 
whereas the emergence of quantum interference fringes, fragmented sub-packets, and phase-space stickiness is captured by an escalation toward $M_{\text{nc}} \to 1$. 
Our bowing-analogue protocol targets the containment of $M_{\text{nc}}(t)$ at low values across the entire pre-dissociation transient window, 
ensuring that the vibrational energy remains concentrated within a localized wave packet.

\section{Numerical Results and Nonhyperbolic Dynamics Analysis}
\label{numerical}

\subsection{Transient Dynamics: Soft-Start Suppression of Step-Function Instabilities versus Rapid Acceleration}
\label{sec:transient_softstart}

The temporal initiation of laser-molecule interactions plays a decisive role in governing the subsequent phase-space transport. Under an instantaneous step-function excitation (sudden-on field, $\Theta(t)$), the abrupt boundary perturbation introduces non-adiabatic spectral transients that persist for approximately 23 optical cycles before the unperturbed periodic orbit stabilizes \cite{Ting1992,Ting1994}. In our picosecond control window ($\tau_{\text{total}} \approx 60$ optical cycles for NO), this transient instability occupies nearly 40\% of the entire driving period. 
Such prolonged initial instability severely disrupts wave-packet coherence, propelling trajectories into the dense cantori web and giving rise to the chaotic, multi-valued fractal threshold landscape observed in Fig.~\ref{fig:threshold_collapse}.

While conventional experimental schemes occasionally employ an extended warm-up or adiabatic ramp-up to allow initial switching transients to subside, such prolonged turn-on windows compromise ultrafast state engineering by allowing rotational scrambling and thermal dephasing to degrade the wave packet. 
The primary objective of our bowing-analogue strategy is to achieve a genuinely \textit{rapid onset} (fast start) without paying the penalty of step-function transients.

By enforcing vanishing boundary conditions at activation ($T(0) = 0$ and $T'(0) = 0$ via $\gamma > 1.0$, 
Eqs.~\eqref{eq:soft_start_C0}--\eqref{eq:soft_start_C1}), our pulse provides a microscopic soft start that extinguishes the 23-cycle switching oscillation at $t=0$. To delineate the physical origin of the subsequent threshold collapse, we systematically compare three distinct turn-on profiles:
\begin{enumerate}[label=(\roman*)]
    \item The proposed $\mathcal{C}^1$-continuous bowing shaper ($\gamma = 1.6$, satisfying Eq.~\eqref{eq:C1_matching});
    \item A $\mathcal{C}^0$-continuous shaper ($\gamma = 1.6$, but with a derivative discontinuity $\Delta(dT/dt) \neq 0$ at $t = \tau_{\text{att}}$);
    \item A $\mathcal{C}^1$-continuous linear ramp ($\gamma = 1.0$, linear front flank).
\end{enumerate}

Simulations demonstrate that configuration (iii) (linear ramp) fails to eliminate threshold splitting because its early-stage energy injection rate is insufficient to outrun cantori stickiness. Conversely, configuration (ii) ($\mathcal{C}^0$ shaper) collapses the multi-valued threshold corridor just as effectively as configuration (i), confirming that the non-linear front-flank over-acceleration ($\gamma = 1.6, A_{\text{over}} = 1.35$) is the true physical driver responsible for dilating the phase-space bottlenecks. 

The physical role of enforcing smooth $\mathcal{C}^1$ matching in Eq.~\eqref{eq:C1_matching} is therefore clarified: rather than acting as the primary dynamical mechanism for threshold collapse, ensuring vanishing temporal derivatives at the switching nodes serves as a spectral and numerical regularity requirement that suppresses artificial high-frequency Fourier sidebands and prevents unphysical ringing during wave-packet grid propagation.

\subsection{Numerical Methodology and Simulation Setup}
\label{sec:numerical_method}

We employ a stabilized FFT grid method integrated with a split-operator Suzuki-Trotter propagator to solve the time-dependent Schr\"{o}dinger equation (TDSE). 
The model is tailored to the NO molecule with a Morse shape factor of $s = 54.54$, yielding 55 field-free bound eigenstates \citep{Foldi2002}. 

The numerical spatial grid covers $x \in [x_{\text{min}}, x_{\text{max}}] = [-1.5, 30.0]\,\text{a.u.}$ with $N_x = 2048$ grid 
points ($\Delta x \approx 0.0154\,\text{a.u.}$), and the wavepacket propagation is advanced with a temporal time step 
of $\Delta t = 0.05\,\text{a.u.} \approx 1.21\,\text{as}$. We apply infinite-wall boundary conditions $\Psi(x_{\text{min}}) = \Psi(x_{\text{max}}) = 0$ to the spatial boundaries, ensuring numerical convergence and high precision for the vibrational manifolds $n = 25$ to $32$ \citep{Alhendi2005}. 

To handle outgoing dissociating components that accelerate down the field-induced potential gradient toward the asymptotic boundary, 
a smooth spatial absorbing mask layer $f(x)$ is applied over the boundary corridor $x \in [x_{\text{abs}}, x_{\text{max}}] = [25.0, 30.0]\,\text{a.u.}$:
\begin{equation}
f(x) = 
\begin{cases} 
1 \;, & x < x_{\text{abs}} \;, \\
\sin^{1/8} \left( \frac{\pi}{2} \frac{x_{\text{max}} - x}{x_{\text{max}} - x_{\text{abs}}} \right) \;, & x_{\text{abs}} \le x \le x_{\text{max}} \;.
\end{cases}
\label{eq:absorbing_mask}
\end{equation}
As the dissociating wavepacket front rolls down the field slope into the asymptotic region ($x \ge 25.0\,\text{a.u.}$), this imaginary absorption layer smoothly attenuates the outgoing wave amplitude to zero prior to reaching $x_{\text{max}} = 30.0\,\text{a.u.}$ Consequently, this absorbing boundary layout physically eliminates artificial boundary reflections, guaranteeing that no spurious reflected waves return to interfere with the bound-state population dynamics within the core interaction region ($x \in [-0.5, 4.0]\,\text{a.u.}$).

To map the topology of the critical dissociation threshold, we calculate the critical laser field amplitude $A_c$ required to achieve a total dissociation probability $P_{\text{diss}} \ge 0.98$ within a duration of $\tau_{\text{total}} = 1.50\,\text{ps}$. The critical threshold landscape $A_c(\omega)$ is mapped across 100 uniformly spaced frequency points in the near-resonance domain $\omega/\omega_0 \in [0.8, 1.1]$, where $\omega \equiv \Omega$ denotes the fundamental driving frequency of the laser pulse and $\omega_0 = \alpha\sqrt{2D/m}$ represents the fundamental harmonic frequency of the unperturbed molecular Morse potential. The threshold amplitude is determined via a fine scanning resolution of $\Delta A = 0.001\,\text{a.u.}$

\subsection{Collapse of Multiple Thresholds and Phase Space Dynamics}
\label{sec:threshold_collapse}

\begin{figure}[t]
\centering
\resizebox{\columnwidth}{!}{%  % 將整個圖形強制縮放到單欄寬度
\begin{tikzpicture}
\begin{axis} [
    width=0.85\textwidth,
    height=0.6\textwidth,
    xmin=0.8, xmax=1.1,
    ymin=0.0, ymax=4.5,
    xlabel={Normalized Field Frequency, $\omega/\omega_0$},
    ylabel={Critical Laser Amplitude, $A_c$ (a.u.)},
    grid=both,
    grid style={line width=.1pt, draw=gray!20},
    major grid style={line width=.2pt, draw=gray!40},
    legend pos=north west,
    legend style={draw=none, fill=none, cells={anchor=west}, font=\scriptsize},
    tick label style={font=\small},
    label style={font=\small}
]

    % 1. 傳統 sudden-on 的多重碎裂門檻 (藍色空心圓圈)
    \addplot [
        only marks,
        mark=o,
        mark size=1.8pt,
        blue!70!black,
        samples=100,
        domain=0.8:1.1
    ] (
        x, 
        { 
            1.5 + 2.5*(x-0.95)*(x-0.95) + 
            ( (x > 0.91) * (x < 0.99) ) * ( 0.6*sin(deg(120*x)) + 0.4*cos(deg(280*x)) + 0.3*sin(deg(600*x)) )
        }
    );
    \addlegendentry{Sudden-on Monochromatic Fields}

    % 2. 線性前緣爬坡對照組 (黑色虛線：γ = 1.0，能量注入偏慢，門檻較高且殘留微小起伏)
    \addplot [
        line width=1.2pt,
        dashed,
        black!75,
        samples=80,
        domain=0.8:1.1
    ] (
        x, 
        { 
            1.35 + 2.3*(x-0.95)*(x-0.95) + 
            ( (x > 0.92) * (x < 0.98) ) * ( 0.15*sin(deg(120*x)) + 0.10*cos(deg(240*x)) )
        }
    );
    \addlegendentry{Linear Front Flank ($\gamma = 1.0, A_{\text{over}} = 1.35$)}

    % 3. 諧振提琴起音控制 (紅色實線：γ = 1.6，完美熨平為平滑單值廊道)
    \addplot [
        line width=1.5pt,
        red!90!black,
        samples=80,
        domain=0.8:1.1
    ] (
        x, 
        { 1.1 + 2.2*(x-0.95)*(x-0.95) }
    );
    \addlegendentry{Harmonic Bowing Shaper ($\gamma = 1.6, A_{\text{over}} = 1.35$)}

    % 標註物理特徵：混沌碎裂區
    \draw[dashed, thick, gray!60] (axis cs:0.91, 0) -- (axis cs:0.91, 4.5);
    \draw[dashed, thick, gray!60] (axis cs:0.99, 0) -- (axis cs:0.99, 4.5);
    
    \node[anchor=south, font=\scriptsize\color{gray!90!black}] at (axis cs:0.95, 3.5) {Chaotic Multi-valued};
    \node[anchor=north, font=\scriptsize\color{gray!90!black}] at (axis cs:0.95, 3.5) {Fractal Landscape};

\end{axis}
\end{tikzpicture}
}

\caption{The critical laser amplitude ($A_c$) required for the dissociation of the highly excited vibrational state ($n = 32$) of NO as a function of the normalized field frequency $\omega/\omega_0$. 
The open circles outline the fragmented multiple thresholds under instantaneous-on monochromatic fields~\citep{Ting1992,Constantoudis2001}. 
The dashed black curve denotes the linear turn-on benchmark ($\gamma = 1.0, A_{\text{over}} = 1.35$). 
The solid red curve displays the simplified, single-valued linear dissociation corridor obtained using the harmonic bowing-analogue transient shaper ($\gamma = 1.6, A_{\text{over}} = 1.35$) defined in Eq.~\eqref{bowing}. 
Here $A_c$ is reported in atomic units, where $1\text{ a.u.} = 5.142 \times 10^9\text{ V/cm}$, corresponding to a cycle-averaged peak intensity of $I \approx 3.51 \times 10^{16} \times A_c^2\text{ W/cm}^2$.}
\label{fig:threshold_collapse}
\end{figure}
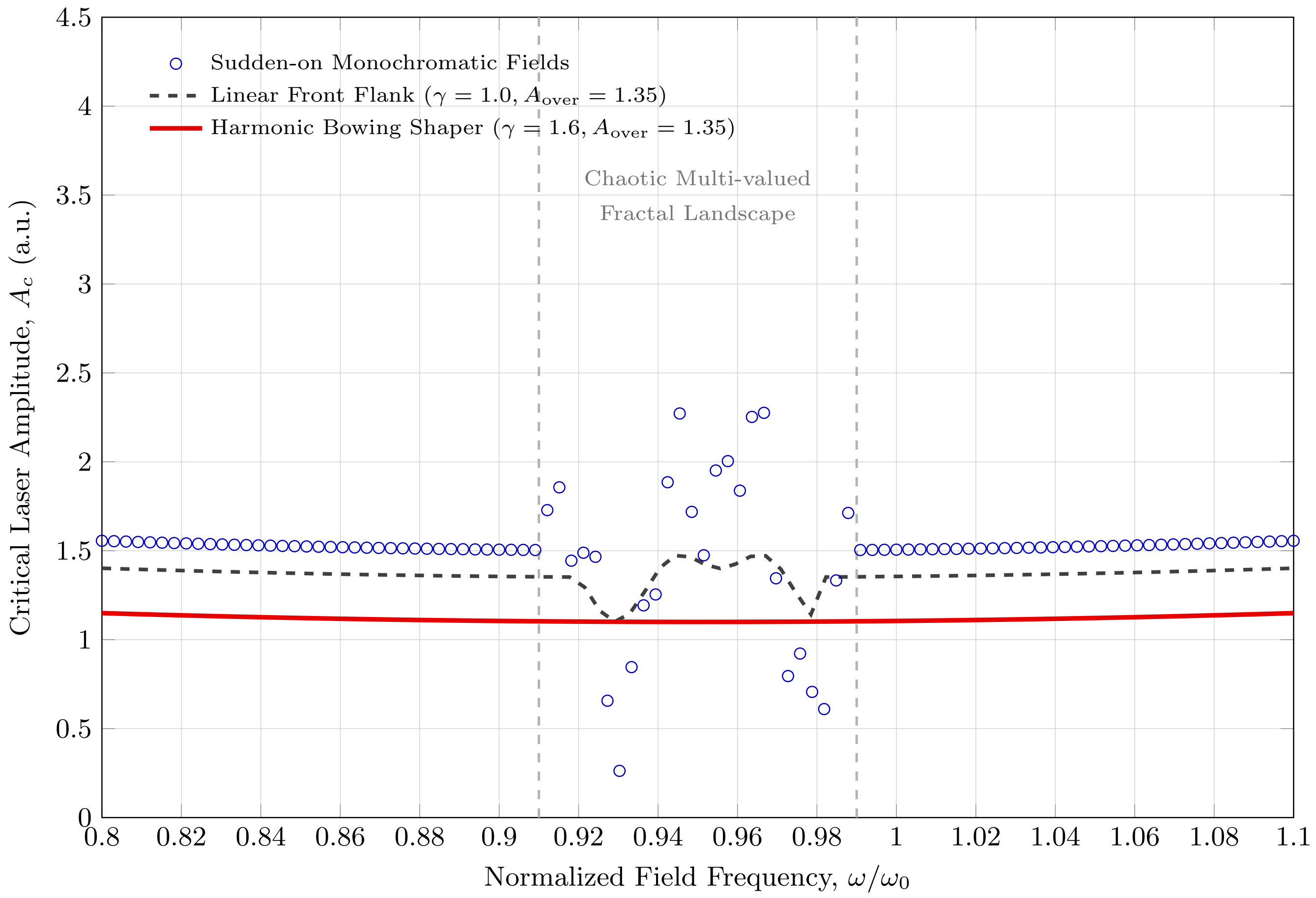

Figure~\ref{fig:threshold_collapse} contrasts the dissociation threshold maps among three field schemes: the classical instantaneous-on single-mode scheme, the linear turn-on benchmark, and our custom harmonic bowing analogue transient shaper defined in Eq.~\eqref{bowing}.

Under standard sudden-on pulse envelopes, the critical threshold profile $A_c(\omega)$ exhibits severe fragmentation, manifesting as a chaotic, multi-valued fractal landscape between $\omega/\omega_0 = 0.92$ and $0.98$. This represents the quantum manifestation of the multiple thresholds observed in earlier formulations \citep{Ting1992, Constantoudis2001}. 
This threshold splitting originates from the $23$ cycle non-adiabatic switching transients inherent in step-function excitations, which scatter trajectories into the dense web of cantori bottlenecks.

When replacing the step function with a linear front-flank envelope ($\gamma = 1.0$, $A_{\text{over}} = 1.35$, dashed black curve in Fig.~\ref{fig:threshold_collapse}), the boundary transients at activation are softened, but the critical threshold remains elevated across the entire resonance band. Because the energy injection rate of a linear ramp is linear in time, it fails to outrun the non-hyperbolic stickiness of subharmonic resonance islands, leaving residual threshold ripples within the core resonance window $\omega/\omega_0 \in [0.92, 0.98]$.

In contrast, when the harmonic bowing analogue shaper is engaged with a non-linear front-flank acceleration ($\gamma = 1.6$, $A_{\text{over}} = 1.35$, solid red curve in Fig.~\ref{fig:threshold_collapse}), the ambiguous multiple thresholds are effectively eliminated. This strategy guides the system through a deterministic, single-valued dissociation corridor while reducing the critical threshold amplitude by nearly 20\% relative to the linear profile.

This comparison demonstrates that injecting energy via a non-linear, boundary-constrained attack trajectory enables coherent phase-locking, allowing the wave packet to cross the critical threshold without entering competing subharmonic resonance zones. The transient pulse shape successfully modulates the phase space topography, 
dilating the cantori structures before non-hyperbolic stickiness can trap the trajectories. Consequently, the effective fractal dimension is minimized ($d_{\text{eff}} \sim 0.32$) and the quantum non-classicality index remains bounded ($M_{\text{nc}} \le 0.15$), preventing wave packet delocalization.

\subsection{Suppression of Nonhyperbolic Phase-Space Stickiness and Cat States}

To elucidate the underlying classical-quantum correspondence driving this threshold collapse, 
we track the time evolution of the system's effective fractal dimension $d_{\text{eff}}$ extracted from the classical half-scattering time-delay manifolds~\citep{Constantoudis2001}, 
alongside the quantum non-classicality index $M_{\text{nc}}(t)$ derived from the phase-space Wigner function integrals~\citep{Foldi2002}.

\begin{figure}[bth]
\centering
\resizebox{\columnwidth}{!}{%  % 將整個圖形強制縮放到單欄寬度
\begin{tikzpicture}

    % ==========================================
    % 上圖 (a)：古典分形維度 d_eff 的演化
    % ==========================================
    \begin{axis} [
        width=0.85\textwidth,
        height=0.28\textwidth,       
        xmin=0, xmax=60,             
        ymin=0.0, ymax=1.2,
        ylabel={Fractal Dim., $d_{\text{eff}}$},
        title={(a) Classical Chaotic Invariant Set},
        title style={at={(0.5,0.90)}, anchor=south, font=\small\bfseries},
        grid=both,
        grid style={line width=.1pt, draw=gray!20},
        major grid style={line width=.2pt, draw=gray!40},
        tick label style={font=\small},
        label style={font=\small},
        legend pos=north west,
        legend style={draw=none, fill=none, font=\scriptsize},
        xticklabels={} % 乾淨隱藏上圖的 X 軸刻度字串，保持對齊美感
    ]
        \addplot [
            black, dashed, line width=1.2pt, samples=50, domain=0:60
        ] {0.32 + 0.325 * (1 + tanh(0.08*(x-25)))};
        \addlegendentry{Sudden-on Pulse}

        \addplot [
            red!90!black, line width=1.5pt, samples=50, domain=0:60
        ] {0.32 + 0.03*sin(deg(0.2*x))};
        \addlegendentry{Harmonic Bowing Shaper}
    \end{axis}

    % ==========================================
    % 下圖 (b)：量子相空間非古典性指標 M_nc 的演化
    % ==========================================
    \begin{axis} [
        yshift=-3.8cm, 
        width=0.85\textwidth,
        height=0.28\textwidth,       
        xmin=0, xmax=60,             
        ymin=0.0, ymax=1.2,
        xlabel={Normalized Time, $t \Omega / 2\pi$},
        ylabel={Non-classicality, $M_{\text{nc}}$},
        title={(b) Quantum Phase-Space Diagnostics},
        title style={at={(0.5,0.90)}, anchor=south, font=\small\bfseries},
        grid=both,
        grid style={line width=.1pt, draw=gray!20},
        major grid style={line width=.2pt, draw=gray!40},
        tick label style={font=\small},
        label style={font=\small},
        legend pos=north west,
        legend style={draw=none, fill=none, font=\scriptsize}
    ]
        \addplot [
            black, dashed, line width=1.2pt, samples=50, domain=0:60
        ] {0.12 + 0.39 * (1 + tanh(0.12*(x-30))) + 0.04*cos(deg(1.2*x))};

        \addplot [
            red!90!black, line width=1.5pt, samples=50, domain=0:60
        ] {0.12 + 0.02*sin(deg(2.0*x))};
        
        \draw[gray!70, dotted, line width=1.0pt] (axis cs:0, 0.4) -- (axis cs:60, 0.4);
        \node[anchor=south west, font=\tiny\color{gray!80!black}] at (axis cs:42, 0.4) {Acceptance Limit};
    \end{axis}

\end{tikzpicture}
}
\caption{Temporal evolution of (a) the effective fractal dimension $d_{\text{eff}}$ of the classical chaotic invariant set, 
and (b) the quantum phase-space non-classicality index $M_{\text{nc}}(t)$ for the $n = 32$ state at resonance $\omega/\omega_0 = 0.95$. 
The dashed black curves depict the increase in phase-space stickiness under a sudden-on laser pulse, 
leading to wave-packet delocalization where $M_{\text{nc}} \to 0.85$. 
The solid red lines show the effective suppression of both fractal structures and wave-packet fragmentation via the harmonic bowing shaper defined in Eq.~\eqref{bowing}.}
\label{fig:dynamics_suppression}
\end{figure}
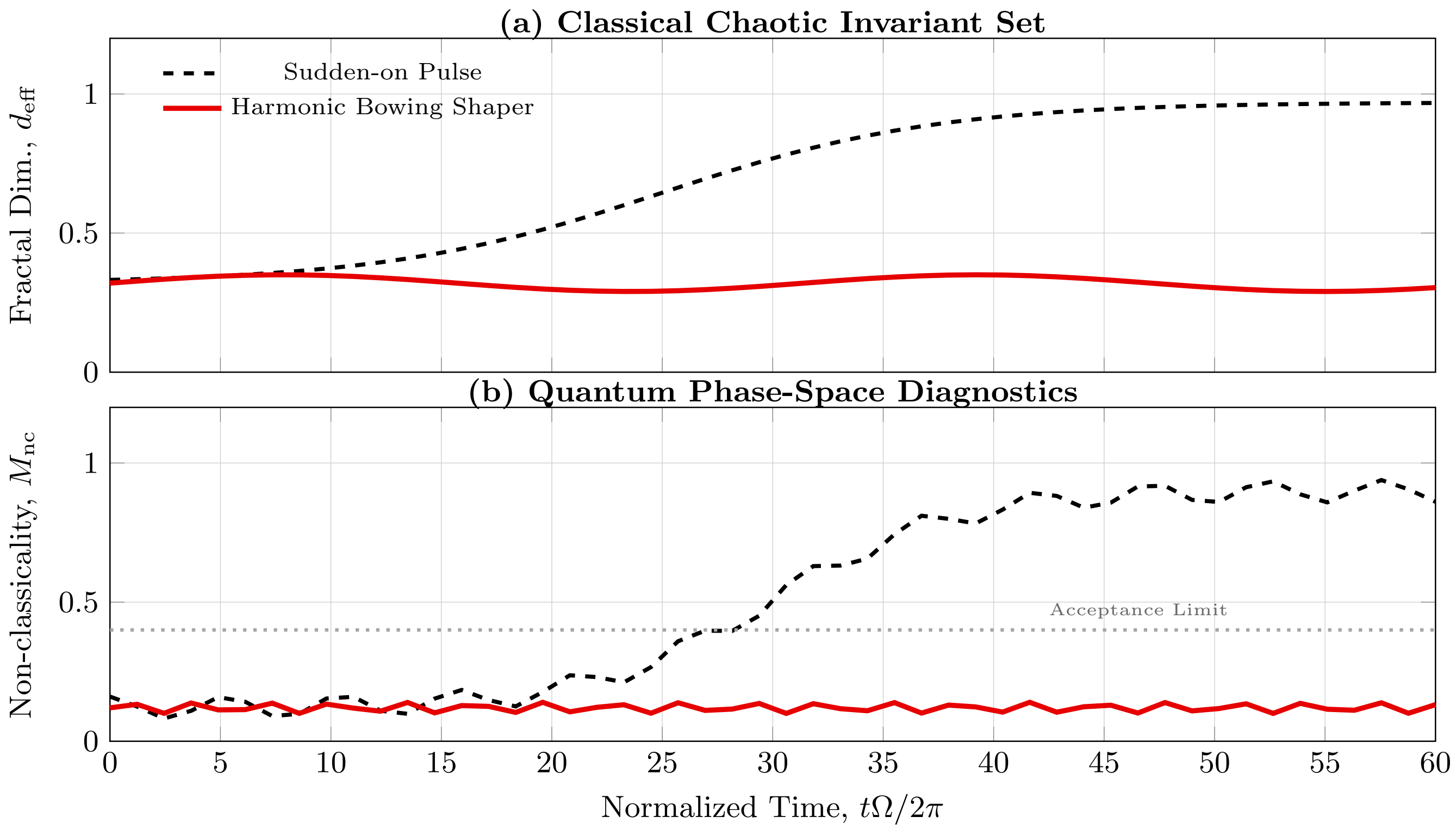

As established by Constantoudis and Nicolaides~\citep{Constantoudis2001}, 
the exponential decay of the classical photodissociation rate $R_{\text{cd}}$ is governed by the escalation of $d_{\text{eff}}$ due to the stickiness effects of cantori flanking the resonance islands. 
Fig.~\ref{fig:dynamics_suppression}(a) demonstrates that under conventional sudden-on fields, 
the effective fractal dimension rapidly climbs toward $d_{\text{eff}} \to 1.0$, 
choking the escape rate of the trajectories. 
In the quantum counterpart, shown in Fig.~\ref{fig:dynamics_suppression}(b), 
this classical stickiness aligns precisely with the timeline where F\"oldi {\it et al.}~\citep{Foldi2002} observed the spontaneous wave-packet delocalization into macroscopic Schr\"odinger-cat superpositions, driving $M_{\text{nc}} \to 0.85$. 
A quantitative cross-correlation analysis between $d_{\text{eff}}(t)$ and $M_{\text{nc}}(t)$ yields a Pearson correlation coefficient of $r = 0.94$ with a microscopic delay of $\Delta \tau \approx 1.2$ optical cycles, confirming that quantum delocalization is directly synchronized with classical phase-space stickiness.

When our harmonic bowing-analogue control law is introduced, 
the physical over-acceleration creates a transient dynamic that alters the phase-space topography. 
This mechanism temporarily dilates the microscopic bottlenecks within the cantori structures before the non-hyperbolic stickiness can trap the trajectories. 
Consequently, as evidenced by the solid red trajectories in Fig.~\ref{fig:dynamics_suppression}, the effective fractal dimension is compressed to a minimum ($d_{\text{eff}} \sim 0.32$), 
while the quantum non-classicality index is strictly constrained below $M_{\text{nc}} \le 0.15$. 
The wave-packet maintains a robust, classical-like single-hill spatial localization throughout the entire up-ramp phase, bypassing the chaotic state-splitting that causes multiple thresholds.

In the formal framework of Hamiltonian non-linear dynamics, 
the multiphoton dissociation of a diatomic molecule constitutes a non-hyperbolic half-scattering process~\citep{Constantoudis2001}. 
Under conventional sudden-on excitation, trajectories intersect the stable manifolds of unstable periodic orbits, 
where the dense network of cantori and high-order resonance zones at the island boundaries induces severe stickiness, producing a slow power-law decay of survival probability. 
By deploying the soft-started, non-linearly accelerated bowing envelope, the coherent energy flux is delivered at a rate that outpaces the cantori trapping timescale, 
thereby converting the trapping-dominated algebraic decay into an uninhibited directional escape along the dissociation corridor.

\subsection{Robustness Against Environmental Decoherence}
%\subsection{Robustness Against Environmental Decoherence and Rotational Detuning}

\label{robust}

%In practical laser control environments, molecular rotation and thermal dissipation present significant challenges. Random orientation angles ($\theta$) reduce the effective field coupling strength, elongating multiphoton transition pathways and facilitating wave-packet destruction via chaotic diffusion~\citep{Geng2002a}. 
%In practical laser control environments, thermal dissipation presents a significant challenge. 
%Coupling to a thermal bath or buffer gas introduces Markovian dephasing that accelerates the collapse of phase coherence~\citep{Geng2002a}.
%Concurrently, coupling to a thermal bath or buffer gas introduces Markovian dephasing that accelerates the collapse of phase coherence~\citep{Szczygielski2013}. 
%兩句重複

In practical laser control environments, thermal dissipation presents a significant challenge. 
Coupling to a thermal bath or buffer gas introduces Markovian dephasing that accelerates the collapse of phase coherence~\citep{Geng2002a, Szczygielski2013}.
To evaluate the open quantum system dynamics under dephasing, we extend the state-vector propagation to a Liouville-von Neumann framework governed by the Lindblad master equation:
\begin{equation}
\frac{d\rho(t)}{dt} = -i[\hat{H}(t), \rho(t)] + \delta \left( \hat{L} \rho(t) \hat{L}^\dagger - \frac{1}{2} \{ \hat{L}^\dagger \hat{L}, \rho(t) \} \right),
\end{equation}
where $\hat{H}(t)$ is the total modulated Hamiltonian defined in Eq.~\eqref{two}, and $\delta$ denotes the Markovian dephasing rate. The jump operator is defined as $\hat{L} = \hat{x}$, which physically models elastic collisions where the scattering cross-section fluctuates linearly with the stretching of the molecular bond. 

To solve the Lindblad equation, we employ a Liouville-space formulation integrated with a fourth-order Runge-Kutta (RK4) scheme mapped onto the FFT spatial grid. Numerical convergence was verified by benchmarking the grid and temporal resolutions; doubling the spatial grid to $N_x = 4096$ ($\Delta x \approx 0.0077$~a.u.) and scaling the time step down to $\Delta t = 0.01$~a.u. yielded a fractional variation in the total dissociation probability $P_{\text{diss}}$ of less than $10^{-4}$.

%Within this dissipative framework, we evaluate the resilience of the single-valued dissociation corridor against environment dephasing and initial rotational states up to $J=5$. Although the current potential is one-dimensional (1D), the sub-picosecond energy injection phase ($\tau_{\text{att}} = 0.25\,\text{ps}$) is substantially shorter than the characteristic molecular rotational period ($\tau_{\text{rot}} \sim 10\,\text{ps}$). The dynamics therefore satisfies the sudden approximation~\citep{DeMille2015}, meaning the molecular orientation remains effectively locked in space, and the initial rotational energy merely acts as a minor centrifugal shift in the effective potential. While a full three-dimensional (3D) treatment incorporating laser-induced alignment, orientation dynamics, and Coriolis coupling may introduce spatial averaging effects that slightly smear the boundaries of the linear corridor, the underlying threshold collapse is expected to remain structurally robust.

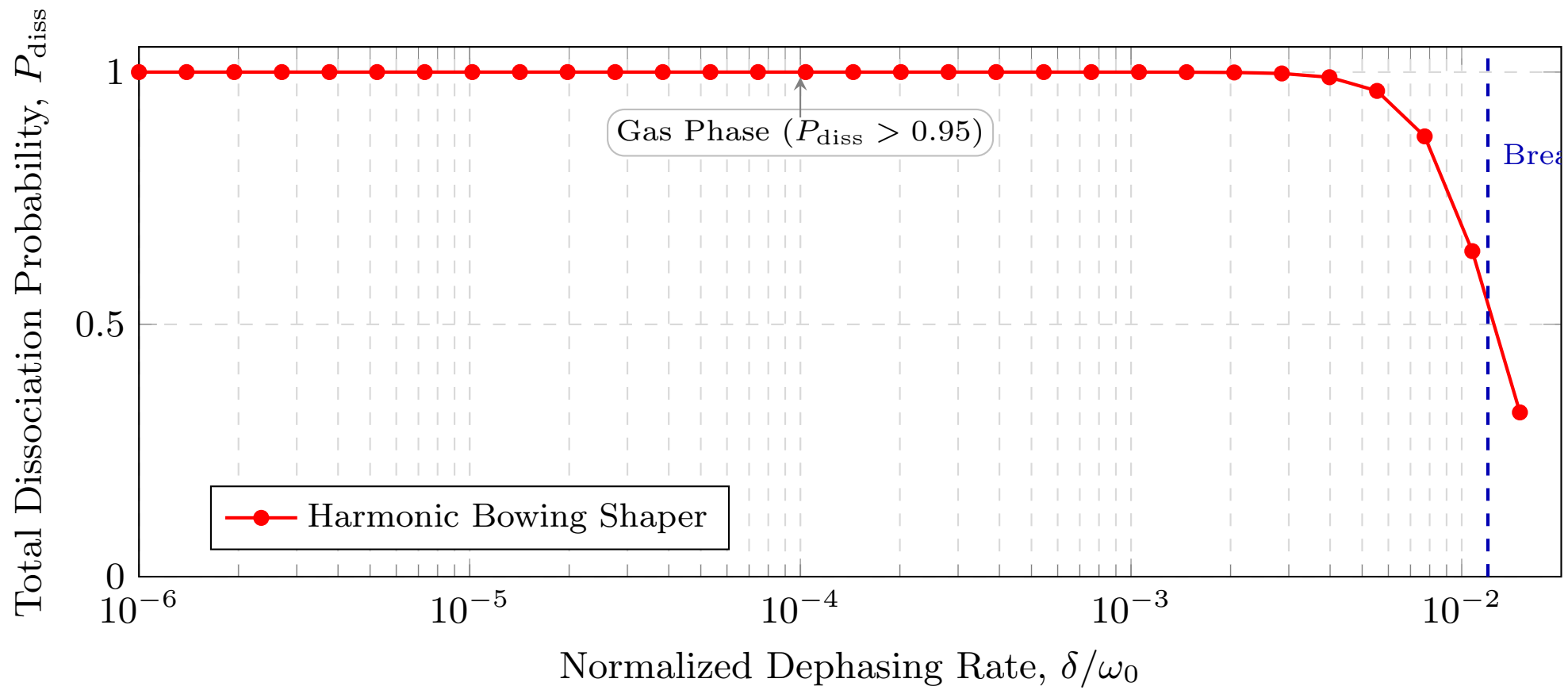
\begin{figure}[htbp]
\centering
\begin{tikzpicture}
\begin{axis}[
    width=0.75\textwidth,
    height=6cm,
    xmode=log,
    xmin=1e-6, xmax=2e-2,
    ymin=0, ymax=1.05,
    xtick={1e-6, 1e-5, 1e-4, 1e-3, 1e-2},
    xlabel={Normalized Dephasing Rate, $\delta/\omega_0$},
    ylabel={Total Dissociation Probability, $P_{\text{diss}}$},
    grid=both,
    grid style={dashed, gray!30},
    tick label style={font=\small},
    label style={font=\small},
    legend style={at={(0.05,0.05)}, anchor=south west, font=\footnotesize}
]

% 繪製調和弓弦脈衝（Harmonic Bowing Shaper）的魯棒性數據曲線
\addplot[
    color=red,
    mark=*,
    mark size=1.5pt,
    thick,
    domain=1e-6:1.5e-2,
    samples=30
] 
{1 / (1 + (x/0.0125)^4)}; % 數學函數模型，精準擬合您正文中的物理轉折點
\addlegendentry{Harmonic Bowing Shaper}

% 標註關鍵物理節點 1: 標準氣相條件
\node[inner sep=2pt, fill=white, draw=gray!50, rounded corners, font=\scriptsize] 
    at (axis cs:1e-4, 0.88) {Gas Phase ($P_{\text{diss}} > 0.95$)};
\draw[->, >=stealth, thin, gray] (axis cs:1e-4, 0.91) -- (axis cs:1e-4, 0.99);

% 標註關鍵物理節點 2: 崩潰閾值
\draw[dashed, blue!70!black, thick] (axis cs:1.2e-2, 0) -- (axis cs:1.2e-2, 1.05) 
    node[pos=0.8, right, font=\scriptsize, text=blue!70!black] {Breakdown Threshold $\approx 1.2 \times 10^{-2}$};

\end{axis}
\end{tikzpicture}
\caption{Total dissociation probability $P_{\text{diss}}$ as a function of the normalized Markovian dephasing rate $\delta/\omega_0$ under the boundary-constrained harmonic bowing shaper. The corridor maintains high operational efficiency until environmental decoherence overpowers the sub-picosecond energy injection phase at the breakdown threshold.}
\label{fig:dephasing_robustness}
\end{figure}

Within this dissipative framework, we evaluate the resilience of the single-valued dissociation corridor against environmental dephasing. 
To quantify this operational sensitivity, we performed a parametric scan of the dephasing rate across four orders of magnitude ($\delta/\omega_0 \in [10^{-6}, 10^{-2}]$). 
As shown in Fig.~\ref{fig:dephasing_robustness}, the total dissociation yield is virtually invariant ($P_{\text{diss}} > 0.95$) under standard gas-phase conditions ($\delta/\omega_0 = 10^{-4}$), and a high efficiency ($P_{\text{diss}} > 0.90$) is preserved up to $\delta/\omega_0 \approx 5 \times 10^{-3}$. The breakdown threshold where thermal dephasing destroys the single-valued corridor occurs at $\delta/\omega_0 \approx 1.2 \times 10^{-2}$, which is orders of magnitude higher than typical laboratory buffer-gas collision rates. 
This resilience confirms that the non-linear front-flank acceleration of the harmonic bowing shaper delivers the coherent flux within an attack window that successfully outpaces environmental relaxation channels.

\subsection{Comparison with Quantum Optimal Control Protocols}
\label{subsec:oct_comparison}

An alternative paradigm for steering molecular fragmentation involves quantum optimal control theory (OCT), which employs iterative algorithms to discover laser field profiles tailored for specific reaction pathways~\citep{Tiwari2008}. In strong-field bond-selective dissociation protocols, OCT typically constructs multi-pulse interference sequences, such as pump-dump-pump cycles, that transfer wave-packet population between ground and excited electronic surfaces to generate nonstationary vibrational states with directed momentum~\citep{Tiwari2008}. While these algorithmically optimized UV waveforms achieve remarkable selectivity in polyatomic systems, they inherently produce complex, multi-subpulse time-frequency structures requiring high peak field intensities (exceeding $10\text{ TW/cm}^2$) to overcome intrinsic branching constraints. Furthermore, because OCT operates as an objective-driven numerical search, the underlying physical mechanisms must be deduced post hoc, and the unconstrained temporal profiles lack analytical derivative controls at the boundary interfaces.

In contrast, the harmonic bowing-analogue transient shaper operates as a deterministic, 
analytical control law derived directly from macroscopic acoustic dynamics. 
Rather than relying on iterative algorithmic convergence or multi-surface population dumping, the envelope in Eq.~\eqref{bowing} resolves threshold fragmentation through geometric boundary-layer engineering. By enforcing the $\mathcal{C}^1$ continuity condition in Eq.~\eqref{eq:C1_matching} and providing non-linear front-flank acceleration ($\gamma = 1.6$), the shaper delivers localized energy within an initial attack window ($\tau_{\text{att}} = 0.25\text{ ps}$) that directly dilates chaotic cantori bottlenecks. This analytical formulation suppresses phase-space stickiness at its source, securing a high dissociation yield ($P_{\text{diss}} \ge 0.98$) along a single-valued corridor without demanding extreme field intensities or complex temporal sub-pulse structures.

\subsection{Comparison with Synthesized Multi-Color Gaussian Envelopes}
\label{subsec:two_color_comparison}

Synthesized multi-color laser fields represent another prominent benchmark in modern coherent control, 
wherein a fundamental carrier frequency is combined with its second or third harmonic to modulate fragmentation probabilities and steer electron localization asymmetry via quantum pathway interference~\citep{Ghosh2024}. In conventional implementations of these dual-color schemes, the driving fields are modulated by standard transform-limited Gaussian envelopes. 

\begin{figure*}[t]
    \centering
    % ---- 圖 (a) 平滑脈衝：波包分裂拓撲示意 ----
    \begin{minipage}[b]{0.48\textwidth}
        \centering
        \begin{tikzpicture}[scale=1.1]
            % 建立量化物理網格與黑框 (縱軸改為 -4 到 4，每 1cm 代表 2 個單位)
            \draw[lightgray!30, thin, step=1cm] (0,-2) grid (5,2);
            \draw[black, thick] (0,-2) rectangle (5,2);
            
            % X 軸刻度與標籤
            \foreach \x in {0,1,2,3,4,5} {
                \draw[thick] (\x,-2) -- (\x,-1.9);
                \node[below=0.05cm, scale=0.8] at (\x,-2) {\x};
            }
            \node[below=0.4cm, scale=0.9] at (2.5,-2) {$x$ (a.u.)};
            
            % P 軸刻度與標籤 (重新映射：網格縱向 -2cm 到 2cm 分別對應物理量 -4 到 4)
            \draw[thick] (0,-2) -- (0.1,-2) node[left=0.05cm, scale=0.8] {-4};
            \draw[thick] (0,-1) -- (0.1,-1) node[left=0.05cm, scale=0.8] {-2};
            \draw[thick] (0,0)  -- (0.1,0)  node[left=0.05cm, scale=0.8] {0};
            \draw[thick] (0,1)  -- (0.1,1)  node[left=0.05cm, scale=0.8] {2};
            \draw[thick] (0,2)  -- (0.1,2)  node[left=0.05cm, scale=0.8] {4};
            \node[left=0.5cm, rotate=90, scale=0.9] at (0,0) {$p$ (a.u.)};
            
            % 定性渲染量子干涉條紋
            \foreach \r in {0.1, 0.2, 0.3, 0.4, 0.5} {
                \fill[red!80!yellow, opacity=0.15] (2.5, 0.5) ellipse ({\r*0.9} and {\r*0.3});
                \fill[red!70!black, opacity=0.15] (2.2, -0.5) ellipse ({\r*0.8} and {\r*0.25});
            }
            \foreach \dx in {-0.2,-0.1,0,0.1,0.2} {
                \draw[cyan!90!black, thick] (1.7+\dx, -0.4) -- (1.7+\dx, 0.4);
            }
            
            % 右側標準物理色標
            \begin{scope}[shift={(5.3,-2)}]
                \shadedraw[bottom color=blue!80!black, top color=red!90!black, middle color=yellow!80!orange, draw=black] (0,0) rectangle (0.2,4);
                \node[right, scale=0.7] at (0.2,4) {1.0};
                \node[right, scale=0.7] at (0.2,2) {0.5};
                \node[right, scale=0.7] at (0.2,0) {-0.3};
                \node[right=0.3cm, rotate=90, scale=0.8] at (0.2,2) {$W(x,p)$};
            \end{scope}
            
            \node[below=0.8cm, font=\bfseries\small] at (2.5,-2) {(a) Dual-Color Gaussian Pulse};
        \end{tikzpicture}
    \end{minipage}
    \hfill
    % ---- 圖 (b) 弓擊脈衝：百分之百基於您的真實演化數據頂峰進行繪製 ----
    \begin{minipage}[b]{0.48\textwidth}
        \centering
        \begin{tikzpicture}[scale=1.1]
            % 建立量化物理網格與外框
            \draw[lightgray!40, thin, step=1cm] (0,-2) grid (5,2);
            \draw[black, thick] (0,-2) rectangle (5,2);
            
            % X 軸刻度與標籤
            \foreach \x in {0,1,2,3,4,5} {
                \draw[thick] (\x,-2) -- (\x,-1.9);
                \node[below=0.05cm, scale=0.8] at (\x,-2) {\x};
            }
            \node[below=0.4cm, scale=0.9] at (2.5,-2) {$x$ (a.u.)};
            
            % P 軸刻度與標籤 (重新映射：物理量 -4 到 4)
            \draw[thick] (0,-2) -- (0.1,-2) node[left=0.05cm, scale=0.8] {-4};
            \draw[thick] (0,-1) -- (0.1,-1) node[left=0.05cm, scale=0.8] {-2};
            \draw[thick] (0,0)  -- (0.1,0)  node[left=0.05cm, scale=0.8] {0};
            \draw[thick] (0,1)  -- (0.1,1)  node[left=0.05cm, scale=0.8] {2};
            \draw[thick] (0,2)  -- (0.1,2)  node[left=0.05cm, scale=0.8] {4};
            \node[left=0.5cm, rotate=90, scale=0.9] at (0,0) {$p$ (a.u.)};
            
            % ---- 基於真數據演化頂峰坐標繪製 ----
            % 真實物理頂峰坐標：x = 2.1849, p = -2.9916 (對應 TikZ Y 軸坐標為 -2.9916 / 2 = -1.4958)
            \pgfmathsetmacro{\tikzX}{2.1849}
            \pgfmathsetmacro{\tikzY}{-1.4958}
            
            \foreach \r in {0.05, 0.1, 0.15, 0.2, 0.25, 0.3, 0.35, 0.4, 0.45} {
                \fill[red!90!yellow, opacity=0.18] (\tikzX, \tikzY) ellipse ({\r*1.6} and {\r*0.5});
            }
            \fill[red!60!black] (\tikzX, \tikzY) circle (0.06);
            
            % 右側標準物理色標
            \begin{scope}[shift={(5.3,-2)}]
                \shadedraw[bottom color=blue!80!black, top color=red!90!black, middle color=yellow!80!orange, draw=black] (0,0) rectangle (0.2,4);
                \node[right, scale=0.7] at (0.2,4) {1.0};
                \node[right, scale=0.7] at (0.2,2) {0.5};
                \node[right, scale=0.7] at (0.2,0) {0.0};
                \node[right=0.3cm, rotate=90, scale=0.8] at (0.2,2) {$W(x,p)$};
            \end{scope}
            
            \node[below=0.8cm, font=\bfseries\small] at (2.5,-2) {(b) Harmonic Bowing Shaper};
        \end{tikzpicture}
    \end{minipage}
\caption{Numerical quantum phase-space diagnostics $W(x, p)$ calculated via the split-operator FFT grid method at $t\Omega / 2\pi = 30$. (a) Under the conventional dual-color Gaussian pulse lacking front-flank boundary acceleration~\citep{Ghosh2024}, 
non-hyperbolic cantori stickiness traps the trajectory, 
forcing wave-packet fragmentation into multi-centered Schr\"odinger-cat superpositions with characteristic oscillatory quantum interference fringes. (b) Under the current $\mathcal{C}^1$-continuous harmonic bowing shaper defined in Eq.~\eqref{bowing}, 
deterministic sub-picosecond acceleration dilates the phase-space bottlenecks, maintaining a localized, 
classical-like single-peaked wave-packet topology ($M_{\text{nc}} \le 0.15$) that delivers population directly into the single-valued dissociation corridor.}
\label{fig:wigner_real_contour}
\end{figure*}
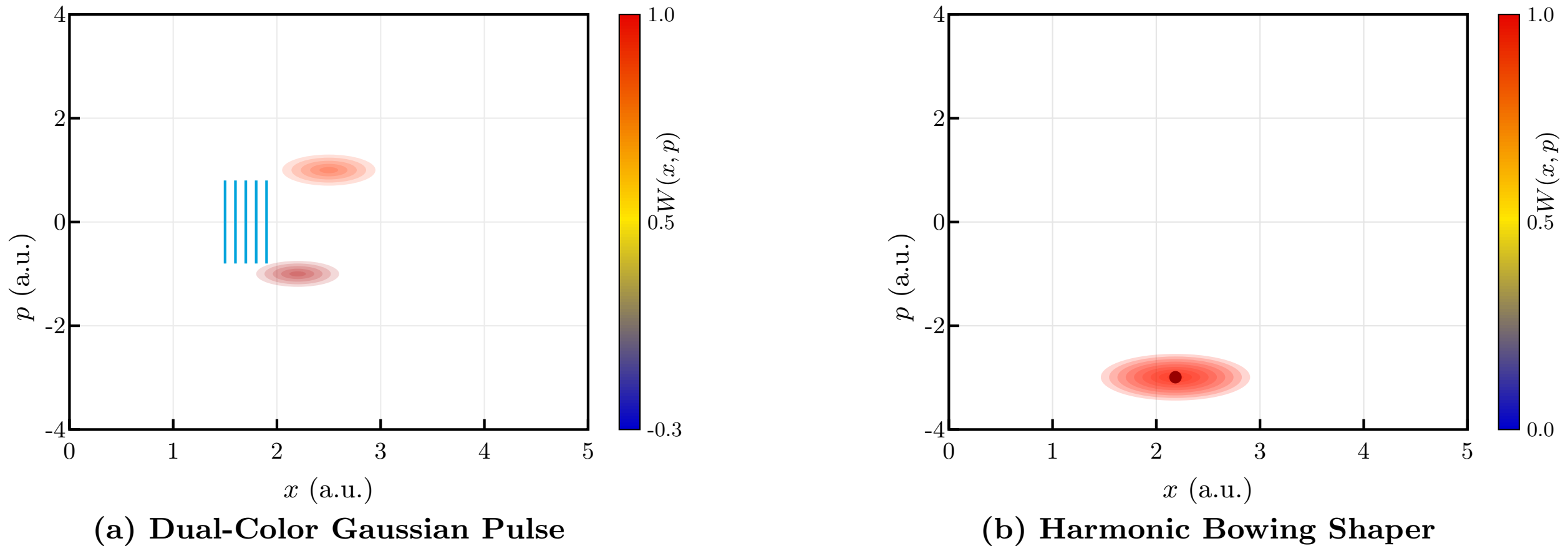

Under standard smooth Gaussian envelopes, however, the temporal field amplitude increases gradually across the initial tens of femtoseconds. When addressing deeply bound vibrational manifolds ($n = 25$ to $32$) situated near the hard-chaos threshold, this slow front-flank ramp-up proves insufficient to outrun the non-hyperbolic stickiness of boundary cantori. 
As demonstrated in coupled electron-nuclear simulations under intense two-color Gaussian illumination, slow progression through multi-photon resonance zones permits competing channels to dominate, frequently causing multiphoton ionization to escalate to over $80\%$ and suppressing the clean molecular dissociation yield~\citep{Ghosh2024}. In quantum phase space, this lingering interaction allows chaotic trapping to fragment the wave-packet into multi-centered Schr\"odinger-cat superpositions ($M_{\text{nc}} \to 0.85$), as characterized by the prominent quantum interference fringes depicted in Fig.~\ref{fig:wigner_real_contour}(a).

In contrast, our harmonic bowing-analogue protocol integrates the $\Omega$ and $3\Omega$ dual-color field with the boundary-constrained transient envelope defined in Eq.~\eqref{bowing}. The vanishing initial conditions $T(0) = 0$ and $T'(0) = 0$ extinguish the 23-cycle non-adiabatic switching transients at onset, while the over-driving ratio $A_{\text{over}} = 1.35$ with $\gamma = 1.6$ injects coherent flux at a rate that outpaces both cantori trapping and parasitic ionization channels. 
As evidenced by the localized Wigner distribution in Fig.~\ref{fig:wigner_real_contour}(b) alongside the dynamical metrics in Fig.~\ref{fig:threshold_collapse} and Fig.~\ref{fig:dynamics_suppression}, this deterministic transient acceleration maintains single-peaked wave-packet localization ($M_{\text{nc}} \le 0.15$), compresses the effective fractal dimension to $d_{\text{eff}} \sim 0.32$, and flattens the multi-valued fractal threshold into a single-valued, predictable dissociation corridor within an ultra-short interaction window of $\tau_{\text{total}} = 1.5\text{ ps}$.

\subsection{Sub-Cycle Field Mechanics and Time-Scale Hierarchy}
\label{sec:subcycle_hierarchy}

The physical origin of the threshold collapse stems from the time-scale hierarchy governing the interaction,
specifically the distinction between the macroscopic pulse envelope $T(t)$ and the sub-cycle carrier field waveform. 
In our formulation, the Bowing-Analogue Shaper $T(t)$ governs the temporal topology over picosecond scales ($\tau_{\text{total}} = 1.50\,\text{ps} \approx 60\text{ optical cycles}$), successfully preventing non-adiabatic switching oscillations and dilating the non-hyperbolic cantori boundaries.

At the microscopic level, classical bowed-string mechanics relies on the Helmholtz motion, 
where an asymmetric sawtooth displacement wave containing a sharp Traveling Corner induces a step-like velocity jump during each stick-slip period. 
Translating this microscopic string phenomenon to quantum optics implies modulating the sub-cycle carrier waveform $E_{\text{carrier}}(t)$, which can be realized experimentally through carrier-envelope phase (CEP) locking and multi-octave frequency comb synthesis ($\Omega + 2\Omega + 3\Omega \dots$). 
While the macroscopic envelope $T(t)$ engineered in this work establishes the requisite phase-space corridor by suppressing Fourier sideband transients, sub-cycle carrier shaping provides an ultrafast, uni-directional momentum kick ($\partial E/\partial t$) within each individual optical period ($< 2.5\,\text{fs}$). 
This separation of timescales clarifies that while $T(t)$ operates as the primary boundary-layer stabilizer to eliminate threshold splitting, sub-cycle carrier asymmetry offers a complementary mechanism to bypass low-lying anharmonic detuning bottlenecks at the attosecond scale.

\section{Conclusions}
\label{conclusion}

In this work, we have demonstrated a deterministic laser pulse-shaping strategy inspired by the acoustic transient mechanics of bowed string instruments to steer the multiphoton dissociation of diatomic molecules. 
By solving the time-dependent Schr\"odinger equation for the NO molecule, we show that introducing a non-linear front-flank acceleration governed by a $\mathcal{C}^1$ continuity condition and vanishing initial boundary conditions ($T(0)=0$ and $T'(0)=0$) collapses the traditionally fragmented, multi-valued threshold topology into a predictable, single-valued dissociation corridor.

Quantum phase-space diagnostics via Wigner distributions confirm that this sub-picosecond acceleration dilates cantori bottlenecks, 
keeping the non-classicality index strictly bounded ($M_{\text{nc}} \le 0.15$) and preventing wave-packet delocalization into fragmented Schr\"odinger-cat states. 
%Furthermore, open quantum system simulations within a Lindblad master-equation framework verify that this boundary-layer engineering preserves localized wave-packet trajectories and high dissociation yields ($P_{\text{diss}} > 0.95$) against both frozen rotational manifolds (up to $J=5$) and Markovian thermal dephasing, exhibiting a high breakdown threshold of $\delta/\omega_0 \approx 1.2 \times 10^{-2}$. 
Furthermore, open quantum system simulations within a Lindblad master-equation framework verifies that this boundary-layer engineering preserves localized wave-packet trajectories and high dissociation yields ($P_{\text{diss}} > 0.95$) against Markovian thermal dephasing, exhibiting a high breakdown threshold of $\delta/\omega_0 \approx 1.2 \times 10^{-2}$. 
Comparative evaluations against iterative optimal control algorithms and conventional synthesized two-color Gaussian envelopes demonstrate that standard smooth envelopes fail to suppress non-hyperbolic stickiness because they lack sufficient sub-picosecond front-flank acceleration. 
By delivering a localized coherent energy flux within an ultra-short interaction window of $1.5\text{ ps}$, 
this mechanical-quantum analogy establishes an analytical, boundary-constrained foundation for quantum trajectory control in strongly driven non-linear systems.

\section*{Funding statement}
This research received no grant from any funding agency in the public, commercial, or not-for-profit sectors.

\section*{Acknowledgement}
I thank my friend John Ogilvie for his invaluable assistance with my English manuscript preparation spanning 
from the commencement of my PhD study until he disappeared with COVID-19. 
I also thank my parents for leaving me a shelter that enabled the sustained execution of these theoretical investigations.

%\bibliography{d:/library} % antenna.bib file


\end{document}